\documentclass{iaus}
\usepackage{graphicx}

\title[Most precise bound to $\Delta\alpha/\alpha$] 
{Most precise single redshift bound to the variability
of the fine-structure constant}

\author[Levshakov et al.]   
{S. A. Levshakov$^1$, M. Centuri\'on$^2$, P. Molaro$^{2,3}$, S. D'Odorico$^4$,
\break 
D. Reimers$^5$, R. Quast$^5$ \and M. Pollmann$^5$}

\affiliation{$^1$Ioffe Physico-Technical Institute, 194021 St.~Petersburg, 
Russia 
\\[\affilskip]
$^2$Osservatorio Astronomico di Trieste,
Via G.~B.~Tiepolo 11, 34131 Trieste, Italy \\[\affilskip]
$^3$Observatoire de Paris 61, avenue de l'Observatoire,
75014 Paris, France \\[\affilskip]
$^4$European Southern Observatory, Karl-Schwarzschild-Strasse 2,
D-85748 Garching bei M\"unchen, Germany \\[\affilskip]
$^5$Hamburger Sternwarte, Gojenbergsweg 112,
21029 Hamburg, Germany 
}

\pubyear{2005}
\volume{232}  
\pagerange{1-2}
\date{?? and in revised form ??}
\jname{Proceedings The Scientific Requirements for Extremely Large Telescopes}
\editors{P. Whitelock, B, Leibundgut \& M. Dennefeld, eds.}
\begin{document}

\maketitle

\begin{abstract}
Verification of theoretical predictions of an oscillating behavior of the
fine-structure constant, $\alpha$, 
with cosmic time requires high precision measurements at
individual redshifts, while in earlier studies the mean
$\Delta\alpha/\alpha$ values
averaged over wide redshift intervals were usually reported.
This requirement can be met via
the Single Ion Differential $\alpha$ Measurement (SIDAM) procedure.
We apply SIDAM to the FeII lines
associated with the damped Ly$\alpha$
system observed at $z = 1.15$ in the spectrum of HE0515--4414.
The weighted mean calculated on base of carefully selected
34 FeII pairs is
$\langle \Delta\alpha/\alpha \rangle =
(-0.07\pm0.84)\times10^{-6}$.
The precision of this estimate represents the absolute improvement with
respect to what has been done in the measurements of
$\Delta\alpha/\alpha$.
\end{abstract}

\firstsection 
\section{Introduction}

The Sommerfeld fine-structure constant,
$\alpha \equiv e^2/\hbar c$, which describes
electromagnetic and optical properties of atoms, is the most suitable
for time variation tests in both laboratory experiments with atomic clocks and
astronomical observations
(for a review see, e.g.,  \cite[Barrow 2005]{B05}).

The question whether or not the 
fine-structure constant 
varied at different cosmological epochs can be answered only through 
observations of quasar absorption-line spectra.
The main requirement of such studies~-- precise line position
measurements at the level of 10$^{-7}$-10$^{-8}$~-- can be 
fulfilled only at giant optical telescopes equipped by
high resolution spectrographs. 

Theoretically the effects of inhomogeneous space and time evolution 
of $\alpha$ were considered by 
\cite{Ma84} and \cite{MB04}.
Most recently \cite{F05} suggested
a damped-oscillation-like behavior of $\alpha$ as a function of
cosmic time $t$.
It is apparent that to
study such irregular changes in $\alpha$,
we need to achieve high precision in the measurements of
$\Delta\alpha/\alpha$ at {\it individual} redshifts,
contrary to the  
averaging procedure over many redshifts
which is usually used to decrease uncertainties of the mean values
$\langle \Delta\alpha/\alpha \rangle$
(\cite[Murphy et al. 2004]{Mu04}, and references therein).

The uncertainties of individual values of
$\Delta\alpha/\alpha = (\alpha_z - \alpha_0)/\alpha_0$,\,
(here $\alpha_0$ and $\alpha_z$
are the values of $\alpha$ at epoch $z=0$ and
at redshift $z$, respectively) 
are currently known at the level of a few ppm (parts per million)
(\cite[Quast et al. 2004]{Q04};  \cite[Chand et al. 2004]{Ch04}).
In both cases the standard many-multiplet (MM) method
(\cite[Dzuba et al. 2002]{Dz02}) has been used.
Further modification of the MM method
(\cite[Levshakov et al. 2005a]{L1})
resulted in a new methodology for probing the cosmological
variability of $\alpha$ on base of pairs of FeII lines
observed in {\it individual exposures}
from a high resolution spectrograph
(henceforth referred to as SIDAM~--
Single Ion Differential $\alpha$ Measurement).

The basic idea behind SIDAM
was to avoid the influence of small spectral shifts due to
ionization inhomogeneities within the absorbers and
due to non-zero offsets between different exposures.
The individual offsets can affect
the shape of the line profiles during rebinning and coadding procedures
when exposures are combined together
to increase signal-to-noise, S/N, ratio (examples are given in
\cite[Levshakov et al. 2005a]{L1}).

\firstsection 
\section{Results}

In our recent paper (\cite[Levshakov et al. 2005b]{L2})
we showed that SIDAM can provide
a sub-ppm precision in a single redshift $\Delta\alpha/\alpha$ measurement
and that this level of accuracy is caused
by {\it unavoidable} intrinsic instrumental imperfections and
systematic errors inherited from the uncertainties of
the wavelength scale calibration. 
We analyzed high quality
spectra of the bright intermediate redshift quasar
HE0515--4414 ($z_{\rm em}$ = 1.73, $B=15.0$; 
\cite[Reimers et al. 1998]{R98}).
The observations were acquired with the
UV-Visual Echelle Spectrograph (UVES)
at the VLT 8.2~m telescope at Paranal, Chile, and the spectral data were
retrieved from the ESO archive.

We found that the weighted mean of the ensemble of $n = 34$ 
$\Delta\alpha/\alpha$ values 
calculated on base of carefully selected FeII lines from
the $z=1.15$ absorber is equal to
$\langle \Delta\alpha/\alpha \rangle = -0.07\pm0.84$ ppm ($1\sigma$ C.L.).
This value is lower than 2~ppm
expected at $z=1.15$ from the damped-oscillatory
model by \cite{F05}.
However, the error of
$\langle \Delta\alpha/\alpha \rangle$ is not small enough to
verify or reject Fujii's model.  
To probe the oscillatory behavior of $\alpha$,
very accurate measurements at higher redshifts are required
where the amplitude of $\Delta\alpha/\alpha$ is expected to be $\sim 5$~ppm. 

The value
$\langle \Delta\alpha/\alpha \rangle~=
-5.7\pm1.1$~ppm found by \cite[Murphy et al. 2004]{Mu04} 
is based on a sample of 143 absorption systems 
observed with the HIRES/Keck spectrograph and
ranging from $z=0.2$ to 4.2.
Now, the higher that 1~ppm accuracy of
$\langle \Delta\alpha/\alpha \rangle$ obtained from the analysis of 
the individual FeII system at $z=1.15$ poses a problem
whether the Keck ensemble average contains some undetected
systematic errors.

As a conclusion, it is worthwhile to note
that the achieved accuracy of $\Delta\alpha/\alpha$ 
is unique for the standard
UVES configuration and that further improvement
at the sub-ppm level can be attained only with
increasing spectral resolution and stabilizing instrumental performance
such as, for instance,
a fiber link producing a stable illumination at the entrance
of the spectrograph and allowing continuous
simultaneous comparison spectrum.

\medskip
S.\,A.\,L. thanks the partial financial support of an IAU travel grant.

\firstsection 


\begin{thebibliography}{}

\bibitem[Barrow 2005]{B05}
     {Barrow, J. D.} 2005,
     \textit{Phil. Trans. Roy. Soc. A} 363, 2139\, (astro-ph/0511440)

\bibitem[Chand et al. 2004]{Ch04}
     {Chand, H., Srianand, R., Petitjean, P. \& Aracil, B.} 2004,
     \textit{A\&A} 417, 853

\bibitem[Dzuba et al. 2002]{Dz02}
     { Dzuba, V. A., Flambaum, V. V., Kozlov, M. G. \& Marchenko, M. V.} 2002,
     \textit{Phys. Rev. A} 66, 022501

\bibitem[Fujii (2005)]{F05}
     {Fujii, Y.} 2005,
     \textit{Phys. Lett. B}  616, 141

\bibitem[Levshakov et al. 2005a]{L1}
     {Levshakov, S. A., Centuri\'on, M., Molaro, P. \& D'Odorico, S.} 2005a,
     \textit{A\&A} 434, 827

\bibitem[Levshakov et al. 2005b]{L2}
     {Levshakov, S. A., Centuri\'on, M., Molaro, P. et al.} 2005b,
     \textit{A\&A} in press\, (astro-ph/0511765)

\bibitem[Marciano (1984)]{Ma84}
     {Marciano, W. J.} 1984,
     \textit{Phys. Rev. Lett.} 52, 489

\bibitem[Mota \& Barrow (2004)]{MB04}
     {Mota, D. F. \& Barrow, J. D.} 2004,
     \textit{MNRAS} 349, 291

\bibitem[Murphy et al. 2004]{Mu04}
     {Murphy, M. T., Flambaum, V. V., Webb, J. K. et al.} 2004,
     in: S. G. Karshenboim \& E. Peik (eds.), 
     \textit{Astrophysics, Clocks and Fundamental Constants},
     (Heidelberg: Springer), p.\ 131 

\bibitem[Quast et al. 2004]{Q04}
     {Quast, R., Reimers, D. \& Levshakov, S. A.} 2004,
     \textit{A\&A} 415, L7

\bibitem[Reimers et al. 1998]{R98}
     {Reimers, D., Hagen, H.-J., Rodriguez-Pascual, P. \& Wisotzki, L.} 1998,
     \textit{A\&A} 334, 96

\end{thebibliography}
\end{document}